# Exotic Electronic Property of Phosphorene Nanoribbons Driven by External Electric Field

Liang Liu, Hongwei Qin and Jifan Hu [a]

School of Physics, State Key Laboratory for Crystal Materials, Shandong University, Jinan 250100, China

By the means of screened exchange density functional theory, we find that the phosphorene nanoribbons with bare zigzag edges that undergo Peierls distortion is a antiferromagnetic semiconductor in which the polarized states are mainly localized at the edges. Under application of external electric fields, the phosphorene nanoribbons present exotic electronic structure transitions including semiconducting, half-metallic, and metallic states.

[a]Author to whom correspondence should be addressed.

Electronic addresses: hujf@sdu.edu.cn and hu-jf@vip.163.com

Graphene is a promising candidate for new electronic device due to its novel electronic properties[1-3]. The gas of massless Dirac fermion in graphene has very high carrier mobility. Usually, graphene lacks a band gap which is required by electronic device to switch on/off the charge flow, only if external electric field is applied to multi-layer graphene to eliminate the spatial inversion symmetry[4]. Such an extra requirement hinders the application of graphene in electronic device. The single layer of $MoS_2$, which is another two dimension (2D) semiconductor and posses a direct gap ~1.9eV and a comparable carrier mobility[5], is more difficult to obtain considering its more complicated chemical gradient. Silicene and germanene are instable at air condition and easily oxidized quickly. As for stanene, it even has not been obtained so far[6-8]. Thus, it is urged to look for other nano material candidates. Black phosphorus is the most stable allotropism of phosphorus with layered structure similar to graphite. Recently, phosphorene, a very thin black phosphorus with even mono atomic layer, has been successfully prepared by a method of chemical exfoliation from black phosphorus[9-11]. Phosphorene, as an analogue of graphene, also posses a high carrier mobility of $1000 cm^2(V\ S)^{-1}$ and a switching (on/off) ratio up to $10^4$. Furthermore, it is an intrinsic semiconductor with a tunable gap (1.51~0.59eV corresponding to number of layers) [9,10,12-14]. Superior to grapheme, phosphorene is expected to be a very promising 2D material as a new generation semiconductor, which is of potential in electronic applications such as field effect transistor (FET).

In order to satisfy the more and more flexible demand of electronic device, magnetic semiconductors are needed to control not only the charge flow but also the spin current. The magnetic semiconductors can be achieved by doping transition metal in semiconductor (forming doped magnetic semiconductor, DMS) or introducing vacancies in lattice ($d^0$ magnetic semiconductor)[15-20]. However, doping of impurity elements into a mono atomic layered material at exact site without the destruction of 2D structure is very difficult to achieve experimentally. However, dangling bonds at edge of 2D structured system may frequently occur, which is expected to induce unpaired spin. If only the magnetic coupling between neighbor spins is strong enough, then the ordered magnetic configuration would be induced[21]. For an extremely important example, the magnetization in graphene zigzag edge would couple with each other in ferromagnetic order at the same side, and the spins at opposite edge would couple with each by anti-ferromagnetic order[21-23]. Furthermore, the edge states at the position of each side with opposite spin polarization would get an energetic opposite translation induced by applying a transverse electric field, due to the Stark effect. Until one special spin tunnel cross the Fermi face, zigzag edge terminated graphene (called zigzag-graphene nanoribbons, ZGNRs) can be tuned to be half-metallic, making it a promising spin-filter device[24,25].

In terms of phosphorene, by standard general gradient approximation (GGA) Zhu et al. has pointed out that the possible edge magnetization would occur at bare or O-saturated phosphorene nanoribbons with zig-zag edge (ZPNRs)[26]. However, after an edge reconstruction (the Peierls distortion), they found a more stable but nonmagnetic state. Du et al. also propose a stable anti-magnetic state at bare ZPNRs by means of hybrid functional (HSE06) calculations[27], but they did not consider the edge reconstruction. The magnetism induced by dangling bonds in this system is still unclear. Moreover, as an analogy to ZGNRs, it is interesting to investigate the influence of an external transverse electronic filed on the edge states in phosphorene.

In this work, based on first-principle calculations, we found that the ZPNRs structure with a stable σ bond connecting the edge atom to its neighbor atoms is energetically favored when Peierls distortion and spin polarization coexist. The application of transverse electric field could induce the transformation from semiconducting to half metallic or metallic states for ZPNRs. These transformations originate from the electrostatic potential divergence between two sides due to Stark Effect. If an applied electric field is strong enough, the band inversion will occur, similar to the Rashiba effect based on spin-orbit-coupling, and the system will show an exotic electronic structure which is similar to quantum spin hall insulator (QSHI).

## Results and Discussion

**Edge reconstruction and spin polarization of bare zigzag phosphorene nanoribbons.** In accordance with previous convention[24], the ZPNRs are classified by number of edge chains in different width as schematized in Fig 1. The width of ZPNRs under study was in the range 0.81 – 2.10 nm, corresponding to N = 4 – 10. As seen in Table 1, with increasing of N (or width), the forming energy per edge atom, $E_{form}$, decreases gradually, which means the structure becomes more stable. One can see the reduction of $E_{form}$ becomes very tiny for N = 8 and 10, revealing the electronic structure tend to be converged. Thus, we focus our attention to the case of N = 8 in the following discussion.

Our work starts on the ZPNRs without edge reconstruction or spin-polarization hypothesis. Two degenerate bands crosses the Fermi level at k = π/a exactly (fig 2a), which is consistent with former reports. Nevertheless, a half filling bands in 1D system will induce a huge instability in the system and the edge structure should undergo Peierls distortion spontaneously. We then consider a double periodicity unit cell to check the edge dimmer reconstruction or possible magnetic structure. The former works[27,28] report the Peierls distortion by GGA, our HSE06 calculations also find both the spin-polarized state and Peierls distortion of edge can avoid the unstable half filling edge state, and induce a gap (Fig 2b, c) by themselves. The optimized geometry structure with Peierls distortion can be seen in Fig 1, the two neighbored phosphorus atoms at each edge do not get close or apart to be dimmer but deviate upwards and downwards in a perpendicular direction to nanoribbons plane, respectively. We mark these two unequal edge atoms as inner atom Pi and outer atom Po (Fig 1). The edge angle $\alpha_1$ is 99.20° for Pi atoms, and $\alpha_2$ is 83.48° for Po atoms. Their bonds with neighbor ions are also slightly different. The lengths of inner bonds are approximate to 2.16 Å, while outer bonds are slightly shorter to 2.14 Å. These divergences between edge atoms shall give exotic electronic structure which will be later discussed in detail. If one only considers the spin polarization for edge atoms, the two possible anti-ferromagnetic (the other magnetic order could not be converged) states are nearly degenerated in energy, and their electronic structures are also very similar.

Now we consider the coexistence of spin polarization and Peierls distortion. Unequal edge atom breaks part of the transition symmetry of system, thus the unequal magnetic structures in our model can be classified into six types (excluding the nonmagnetic structure). in convention to former work[29], we mark the spin up and down state by '+', '-' sign, and arrange the sign in order of (P1i, P1o, P2i, P2o), as shown in Fig.1. The magnetic order are noted as FM(+ + + +), the rest antiferromagnetic arrangements are labeled as AFM1(+ − + −), AFM2(+ − − +) and AFM3(+ + −

−), which are shown in Fig 3. Note that the two mixed configurations of MIX1(− + + +) and MIX2(+ − + +) are not needed to be considered, since they are unstable.

**Magnetic ground state of bare zigzag phosphorene nanoribbons and its mechanism.** In our calculations, only in narrow stripe model (N = 4, 6) can we get a converged FM and AFM3 state. While N is larger than eight, both initial hypothetical configurations of FM and AFM3 are finally converged to AFM1 state. The spin polarized states mainly locate in edge atoms, and the tails of edge states spread into bone core more or less. The density of states (DOS) projected to edge atoms and atoms near edge are plotted in Fig 4. Due to the folding structure of phosphorene, the bone atoms can be divided into two types, the ones are located at the same plane with edge atom approximately, while the other ones are not. In Fig 4, the first one type are marked as 'u' and the second one as 'd'. These projected DOS show the evident spin splitting, and the energy gap of minority spin states becomes smaller than majority spin states. The tails of edge states prefer to spread on 'u' layer atoms even if some 'd' layer atoms are closed to edge. Furthermore, it should be noted that the tails of edge state localized at each bone atoms keep the same spin polarized direction with its nearest edge atom. This feature is contrast to ZGNRs, in which the tails of edge states always keep the anti-paralleled spin polarized direction to nearest site[22].

The relative energies of four possible magnetic configurations are listed in Table 2. All of them are lower than that of nonmagnetic state, which contradicts with former theoretical work based on GGA-PBE[26], in which they give a nonmagnetic ground state. Note that we also get a nonmagnetic ground state on Peierls distorted stripe by means of GGA-PBE, only by HSE06 calculations can we get the magnetic ground state. From Table 2, we can see that the relative energies of each AFM states is lower than FM state, indicating that the anti-paralleled spin exchange interaction between phosphorus atoms is favored. Thus the exchange interaction between magnetic moments along each edge should be anti-ferromagnetic, which is contrast to ZGNRs[21,22]. Furthermore, we can also see from Table 2 that the interspins chains trend to be anti-parallel too. It can be understood by the tails of edge states. Fig.5 shows the contour plots of spin densities in a schematic model of N = 4 narrow strip, where the different color represents the different spin directions. Due to the features of edge-state-tails discussed above, for the AFM1 state, the edge-state-tails derived from the opposite edge should make the spins meeting at the center of stripe be anti-parallel. As for AFM2 state, the edge-state-tails make the spins meeting at center of stripe be parallel. Thus, AFM1 state is energetically favored due to the anti-ferromagnetic exchange energy. From the projected DOS in Fig 4, one can see that the tails of edge states decays rapidly. This indicates that ZPNRs have a more localized edge states than ZGNRs. Therefore, when the width of stripe is large enough, edge-state-tails meeting in center of stripe almost become negligible, and the energy divergence between AFM1 and AFM2 state almost disappears as see in Table 2.

**Influence of symmetry broken by Peierls distortion on electronic structure.** Although the energy between AFM1 and AFM2 states are nearly degenerated when N is larger than 8 in our model, structure divergence between their bands is obvious, as seen in Fig 6a and 6c. The spin resolved band structure of AFM1 state shows spin degeneration, while the AFM2 state does not. It originates from the characters of distribution of edge states and the symmetry of the stripe. As discussed above, after Peierls distortion the system loses a part of transition symmetry but keeps centrosymmetry. For instance, edge atom $P_{1i}$ and $P_{2i}$ are exactly symmetric about the center of mass, and so are $P_{1o}$ and $P_{2o}$. Thus, the edge states localized on $P_{1i}$ and $P_{2i}$ should be energetically degenerated, and so are $P_{1o}$ and $P_{2o}$. By our calculations, we found that the energy level of outer edge atom ($P_{1o}$, $P_{2o}$) is never lower than that of inner atom ($P_{1i}$, $P_{2i}$). In the right region of Fig 6, we plot the configuration of corresponding edge states in real space.

In terms of k = π/a, each edge state is strongly localized on edge atoms of the same side, and the divergence of distributions on two neighbor edge atoms is robust. In the AFM1 magnetic structure, the spin configuration has broken the centrosymmetry, the edge state localized on $P_{1i}$ is a spin-up state, while the edge state localized on $P_{2i}$ is a spin-down state, as shown in Fig 3(c). This pair of opposite spin states form one of the spin degenerated point at lower position in Fig 6(a). Meanwhile, the other pair of opposite spin states form the spin degenerated point at higher position. For the AFM2 magnetic structure, as seen in Fig 3(d), the spin configuration has kept the centrosymmetry exactly. Both of the edge states localized on $P_{1i}$ and $P_{2i}$ are spin-up state, forming the spin-up energetic degenerated point. And both of the edge states localized on $P_{1o}$ and $P_{2o}$ are spin-down state, forming the spin-down energetic degenerated point.

**Semiconducting, metallic and half metallic transitions driven by transverse electric field.** It has been widely accepted for GNRs that band gaps can be tuned by applying external transverse electric field. It is interesting to investigate the influence of transverse electric field upon the novel 2D semiconductor ZPNRs. The band evolution of ZPNRs with respect of transverse electric field can be seen in Fig 7. Transverse electric filed broke the spatial inversion symmetry of nanoribbons, so that the degeneracy induced by centrosymmetry disappeared in all cases. Furthermore, due to the Stark effect, the edge states, which are well localized on edge atoms, should get an extra electrostatic potential and shift to different direction according to the relative position of each state in electric field.

For the AFM2 magnetic configuration, we plots the variation of energy gap in the two opposite spin states related to electric field in Fig 5. Under a slight electric field, one can see an approximately linear variation of gap related to electric field, corresponding to the linear variation of electrostatic potential. When the external electric filed is beyond critical value of ~0.2V/Å, the top valence band and bottom conduction band crosses each other near k = π/a. The energy gap in minority spin state is totally eliminated, however for the majority spin state, energy gap still keeps a finite value. This shows a half-metallic behavior which can induce a completely spin polarized electrical current. This property may be applied in spin filter device[24,29]. For the other anti-ferromagnetic configuration AFM1, the total energy gap evolution is similar to former. The difference is that the opposite spin band will meets at a critical electric field (~0.2V/Å). As stronger electric field is applied, band reversion near k= π/a occurs too. In this case, although half-metallic state cannot be achieved, the reversed band structure is even more exotic since it is very similar to the edge bands in quantum spin hall insulator[30-32]. The spin tunnels are fixed to momentum. However the band reversion in our model is essentially mediated by external electric field, not the spin-orbital coupling effect in quantum spin hall insulator.

In conclusion, we have investigated the property of bare ZPNRs in detail. The stripe is stable while edge reconstruction (Peierls distortion) and spin polarization coexist. Contrast to ZGNRs antiferromagnetic order is energetically favored at each side of ZPNRs. We propose the origin by analyzing the characteristics of edge states. Furthermore, the Peierls distortion breaks part of the transition symmetry of the system thus breaks the energetic degeneration of states which are localized in neighbor edge atom. We find an electronic structure transition under the external transverse electric field for the different magnetic configurations. One of the antiferromagnetic configurations can be tuned into half metallic, which may have some application in spin filter devices. The other can be tuned into a more exotic electronic structure which is similar to quantum spin hall effect.

## Method

All our calculations have been performed within the density functional theory (DFT) formalism using PAW method as implemented in Vienna ab initio simulation package (VASP) code package[33,34]. An energy cutoff of 400 eV was adopted for the plane-wave expansion of the electronic wave function. Brillouin zone was sampled to a (5×1×1) meshes. The energy convergence criteria was set to $10^{-5}$ eV, and the position of each atom is allowed to relaxation until the max residual force is less than 0.01eV/A. We have employed the Heyd-Scuseria-Ernzerhof hybrid-functional, which incorporates about ten percent nonlocal Hartree-Fock exchange energy in the short range electron-electron interaction region[35-38]. This method has been shown to be superior to local density approximation or general gradient approximation in electronic structure description, especially the prediction of correct qualitative band gap in small gap systems, and give a reliable results in some similar nano systems[25,29,39]. In order to simulate the external electric field in a ZPNRs, a periodic saw-tooth-type potential perpendicular to the direction of the ribbon edge is used (Fig 1).

## Acknowledgments
We acknowledge support by National Natural Science Foundation of China (Grant Nos. 51472150,

51472145 and 51272133), Shandong Natural Science Foundation (Grant No. ZR2013EMM016),

and National 111 Project (B13029).

## Author contributions
Jifan Hu and Hongwei Qin proposed the main ideal of this work. Liang Liu did numerical calculations. Liang Liu and Jifan Hu wrote the manuscript. All authors commented on this work.

## Additional information
**Competing financial interests:** The authors declare no competing financial interests.
11

**Figure Legends:**

Fig 1  Diagram of ZNPRs (N=8 here), and sketch of external transverse electric field. The nanoribbons were assumed to be periodic along x direction. The left region between the two dashed lines denotes a unit cell while edges undergo Peierls distortion. Right one is the side view shows how the two neighbor edge atoms stagers each other in z direction. The grey arrow note the electric field is along the negative y direction.

Fig 2  (a) The band structure of ZPNRs without the consideration of spin polarization and Peiers distortion; (b) The band structure of spin polarized ZPNRs without Peiers distortion. The bands resolved to majority and minority spin are coincide exactly; (c) The band structure of spin unpolarized ZPNRs with Peiers distortion. In all cases, the Fermi energy noted by dashed line is set to zero.

Fig 3  Schemes of four unequal magnetic configurations under consideration (a) FM, (b) AFM3, (c) AFM1 and (d) AFM2, respectively. The angles represent phosphorus atoms, thin lines represent the bonds. Blue and red arrows which point different direction indicate the sign of edge magnetic moment. Due to the inequality of neighbor edge atoms, the magnetic moments of neighbor edge atoms are classified into two types. Solid and dotted arrows represent the magnetic moments for outer and inner edge atoms respectively.

Fig 4  DOS projected to each atom. Each spin resolved DOS are separated into blue, positive part and red, negative part, indicating the majority and minority spin states respectively. The zero point at each energy axis is shifted to Fermi energy. In the right up region, the side and top view of a part of strip which is close to edge are shown. Position of each atom is noted by u0, u0', u1, u1', dn, dn' represent the two edge atoms, the two bone atoms in 'u' layer and the two bone atoms in 'd' layer respectively. These notes are also shown at the bottom of each projected DOS inset to indicate the corresponding atom.

Fig 5 Counter plots of spin densities of AFM1 and AFM2 states in a narrow strip N = 4 model. The thick purple lines plot top view of the ZPNRs wireframe model, where the positions of each critical atom is noted to clarify, the definitions of these notes are given in main text.

The red and blue circles represent the spin up and spin down densities respectively. The counter plots are got within the plane which is paralleled to the stripe and contains the center of model.

Fig 6  The electronic structure of ZPNRs in AFM1 state under external electric field of (a) 0; (b) 0.3 V/Å respectively. The electronic structure of ZPNRs in AFM2 state under external electric field of (c) 0 and (d) 0.3 V/Å respectively. In all cases, left region plots the spin resolved band structure, blue and red lines denote bands of majority and minority spin respectively, and dashed lines denote the Fermi energy which is set to zero. Right region plots the isosurface of band decomposed charge density at momentum k = π/a, which are marked in band structure by grey dots. The isosurface level is set to $10^{-3}$ Å$^{-3}$. In (a), (c), the bands at k = π/a are double degenerated, thus we plot two corresponding isosurface of charge density for each grey dot. In (b), (d), the grey arrows at middle of charge density denote the direction of external electric field. Under application of a transverse external electric field in the direction from the left to the right, the electrostatic potential of right edge increases, however, the one of left edge decreases. Therefore, the energy levels of right edge states are raised and those of left edge states are lowed.

Fig 7  Evolution of band structure of ZNPRs: (a) AFM1 state and (b) AFM2 state. In all cases, the Fermi levels are set to zero and noted by dashed lines. Blue and red solid lines denote the bands of majority and minority spin respectively. From up to down, the external transverse electric filed is 0.1, 0.2, 0.25, 0.3 V/Å respectively. Under an electric filed of ~0.2 V/Å, the energy gap is eliminated. Then, the AFM1 state gets into metallic state while AFM2 state gets into half-metallic state since Fermi level just crosses the bands of minority spin, and the majority spin still keeps a gap.

Fig 8  Variation of energy gap for majority spin (blue line) and minority spin (red line) of ZNPRs in AFM2 state as a function of external electric field.

Table 1  The forming energy for each ZPNRs with different N corresponding to the width.

| N | $E_{form}$ (meV/ edge atom) |
|---|---|
| 4 | 872.18 |
| 6 | 862.88 |
| 8 | 814.00 |
| 10 | 817.93 |

Table 2  The relative energy for different magnetic configuration of different width of ZPNRs, with the energy of nonmagnetic state as reference.

| | Relative Energy (meV/ edge atom) | | | |
|---|---|---|---|---|
| N | + + + + | + + − − | + − − + | + − + − |
| 4 | -49.15 | -38.98 | -69.25 | -57.88 |
| 6 | -5.85 | 30.33 | -43.13 | -43.00 |
| 8 | - | - | -18.88 | -18.70 |
| 10 | - | - | -18.05 | -18.03 |

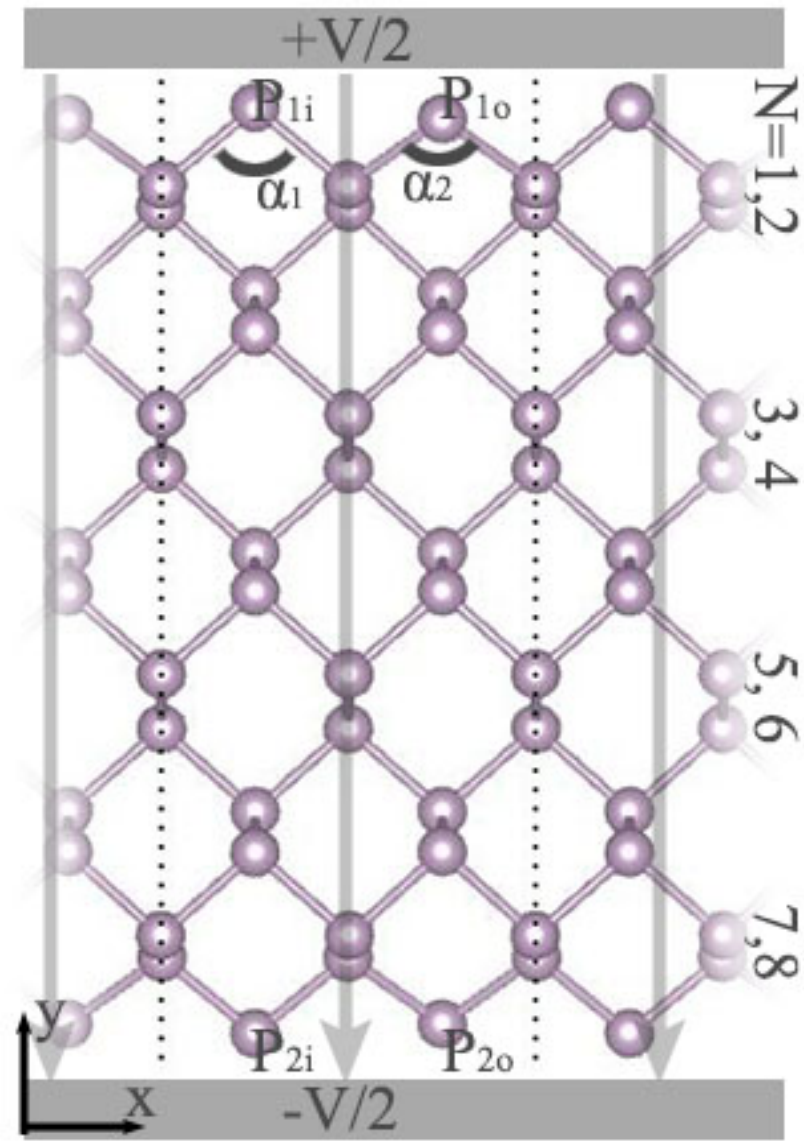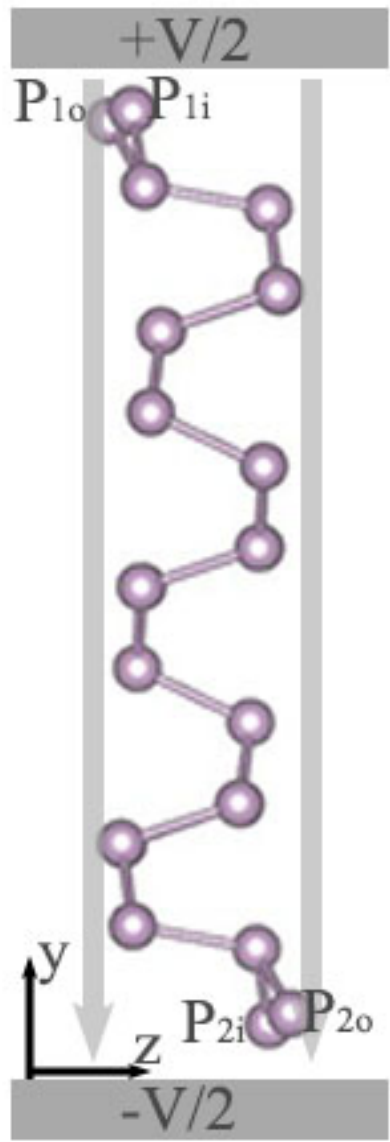

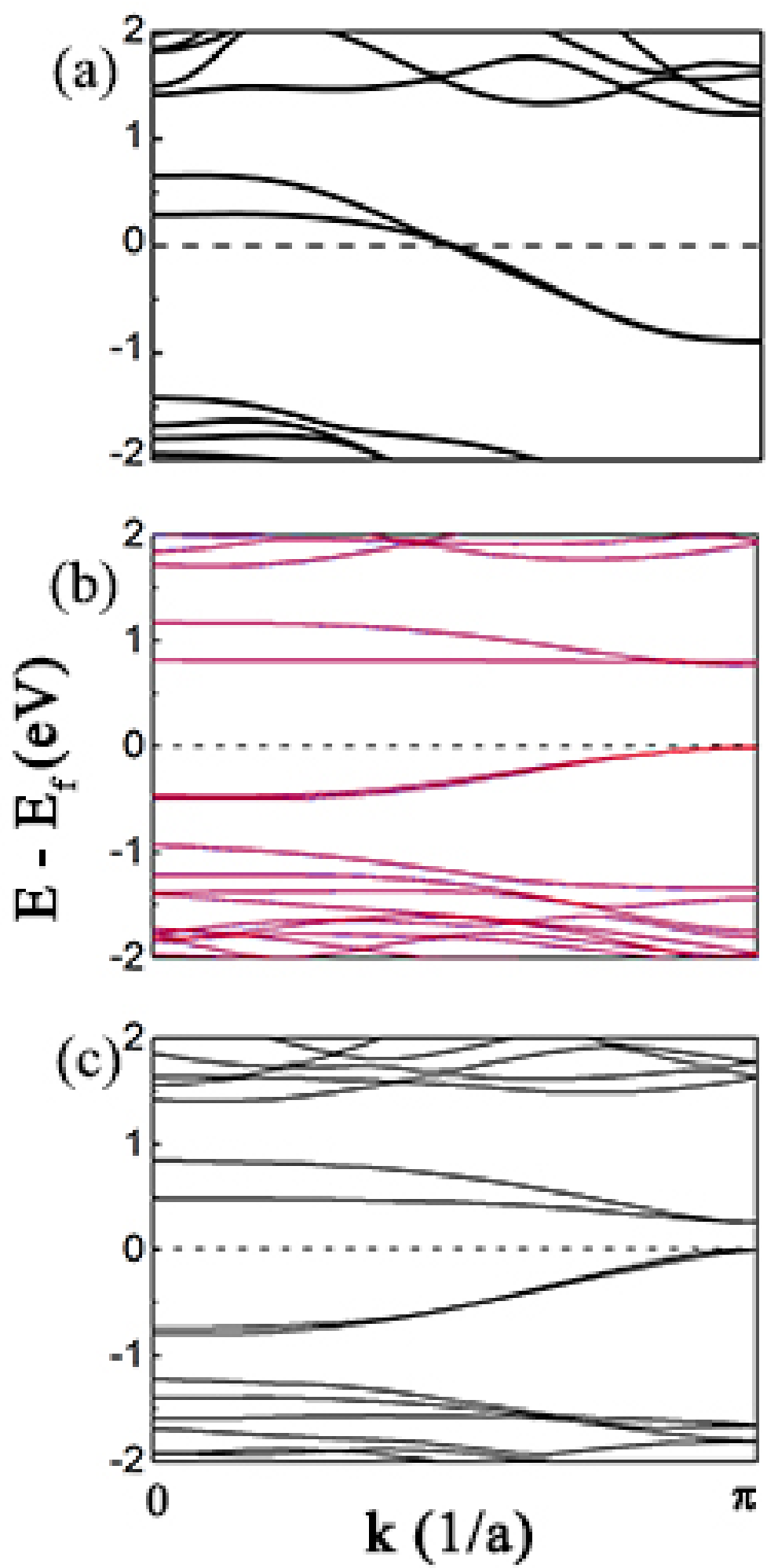

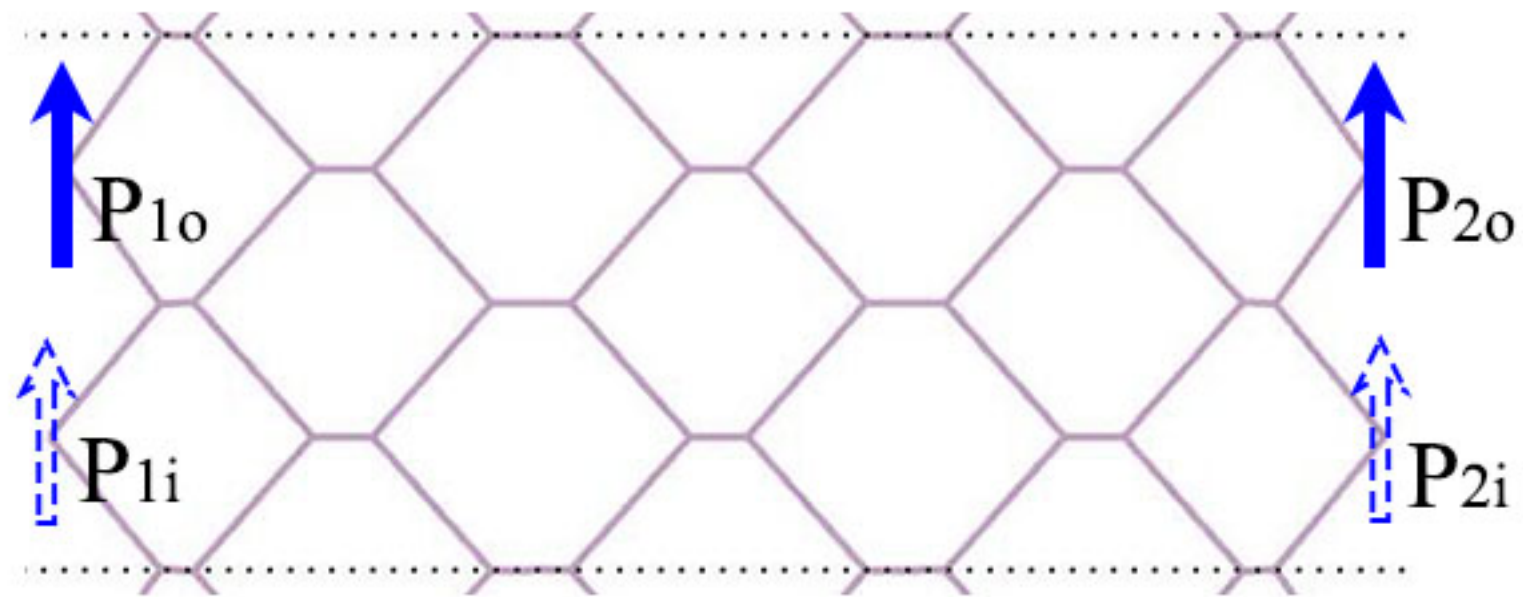 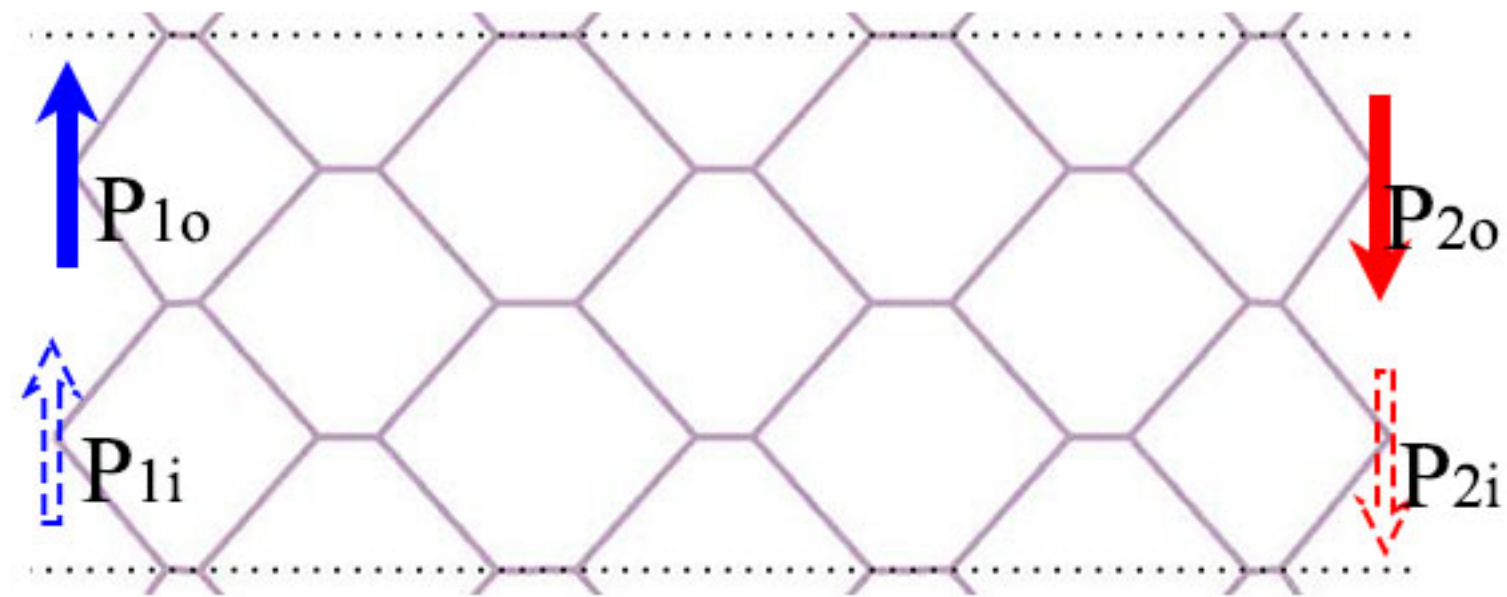

(a) FM(++++)     (b) AFM3(++--)

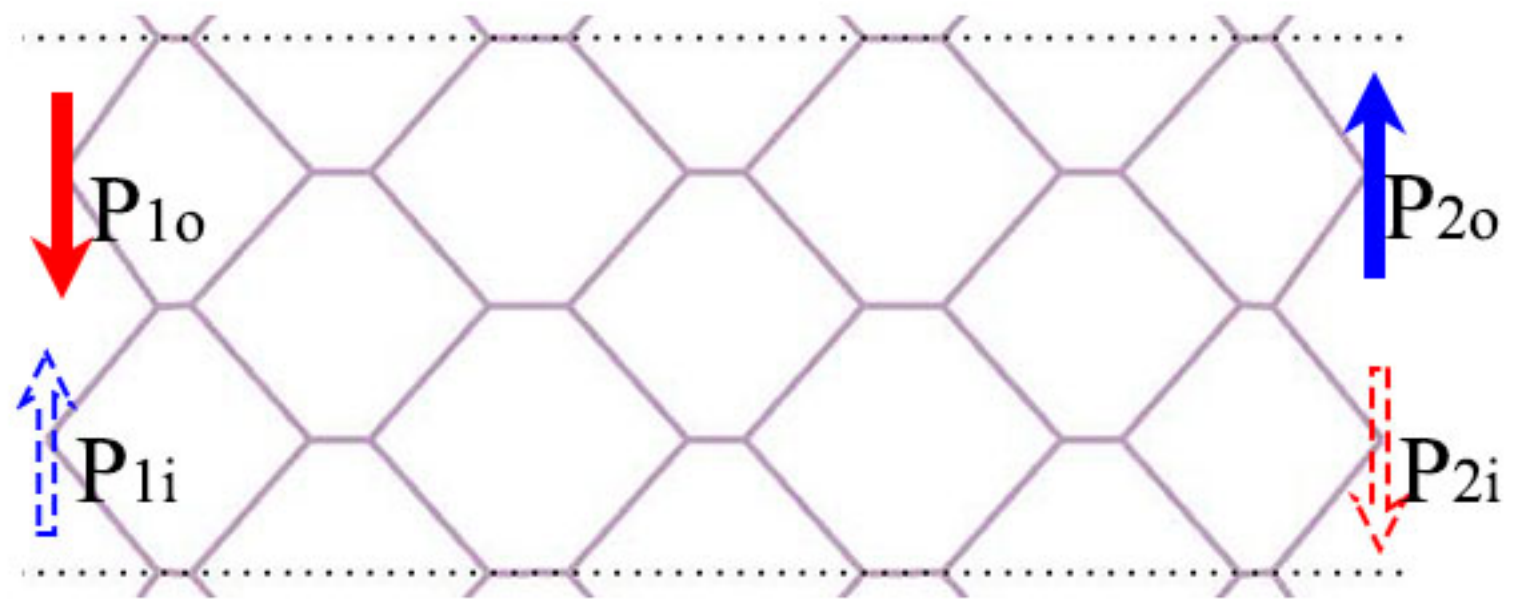 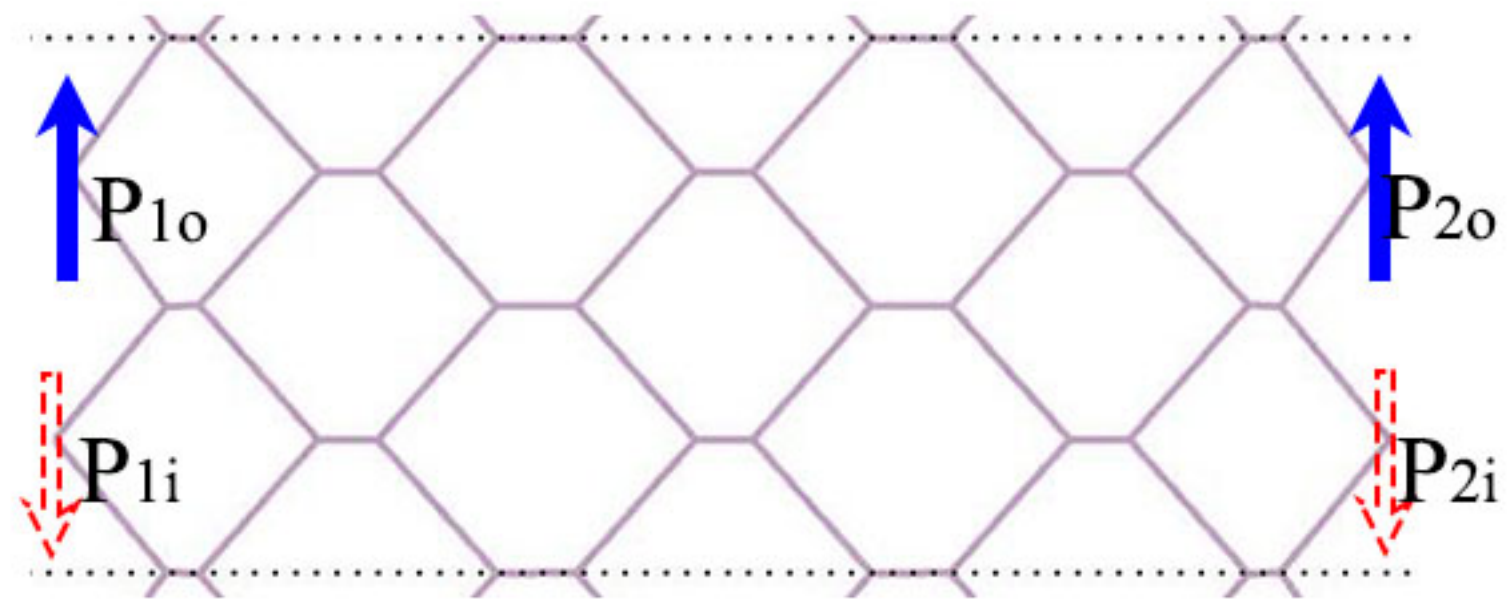

(c) AFM1(+--+)     (d) AFM2(+-+-)

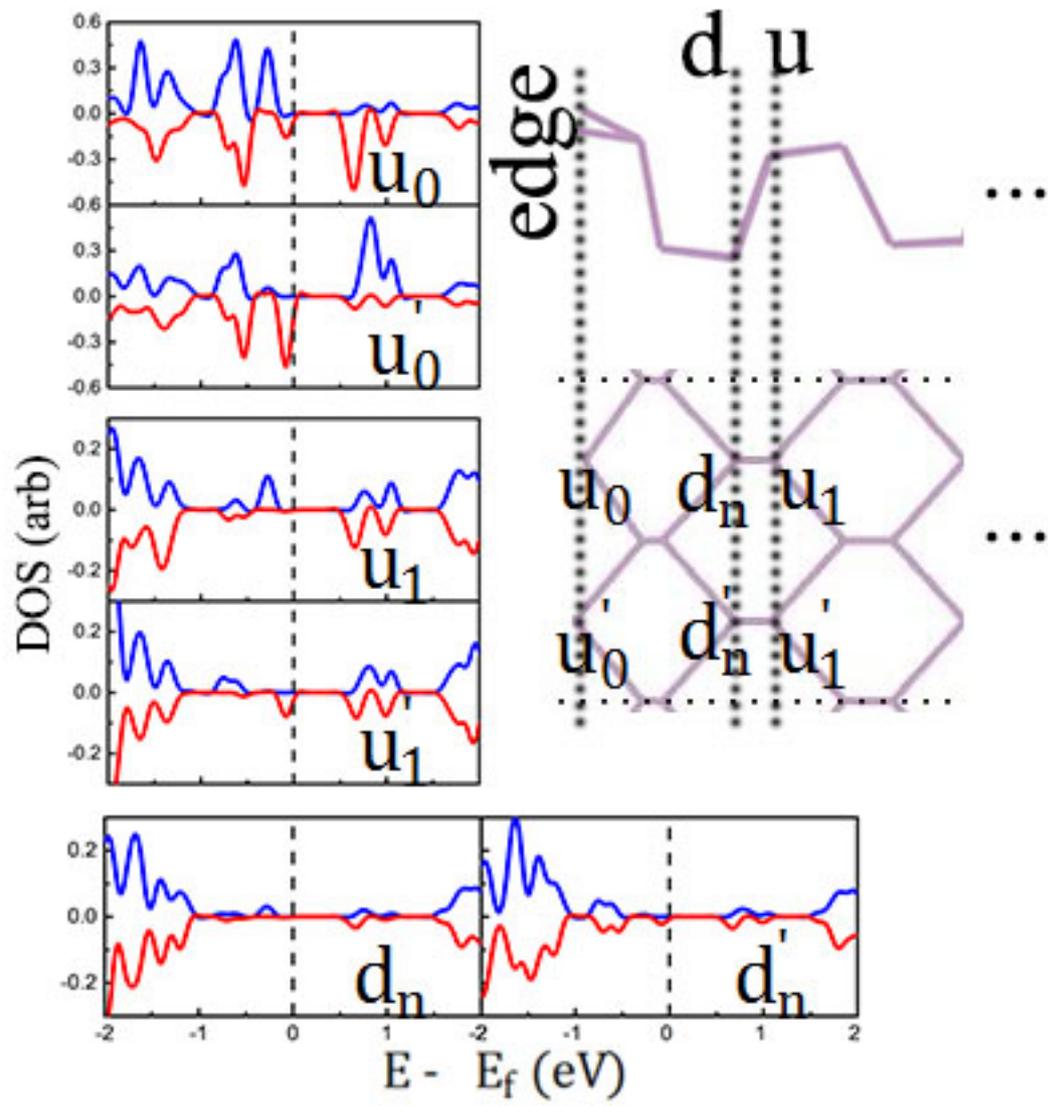

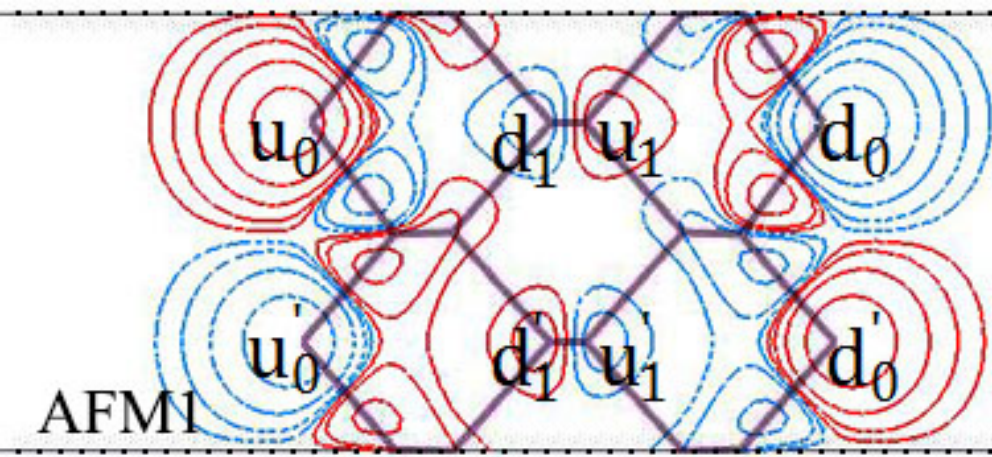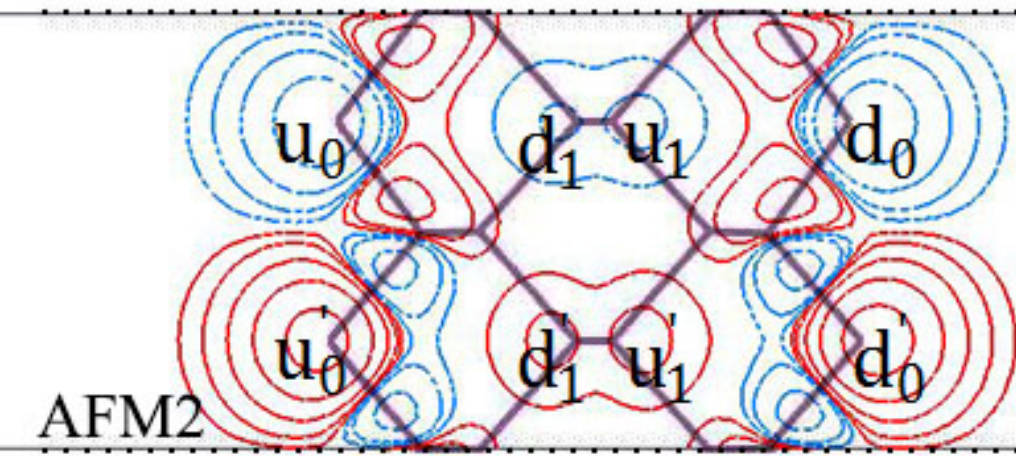

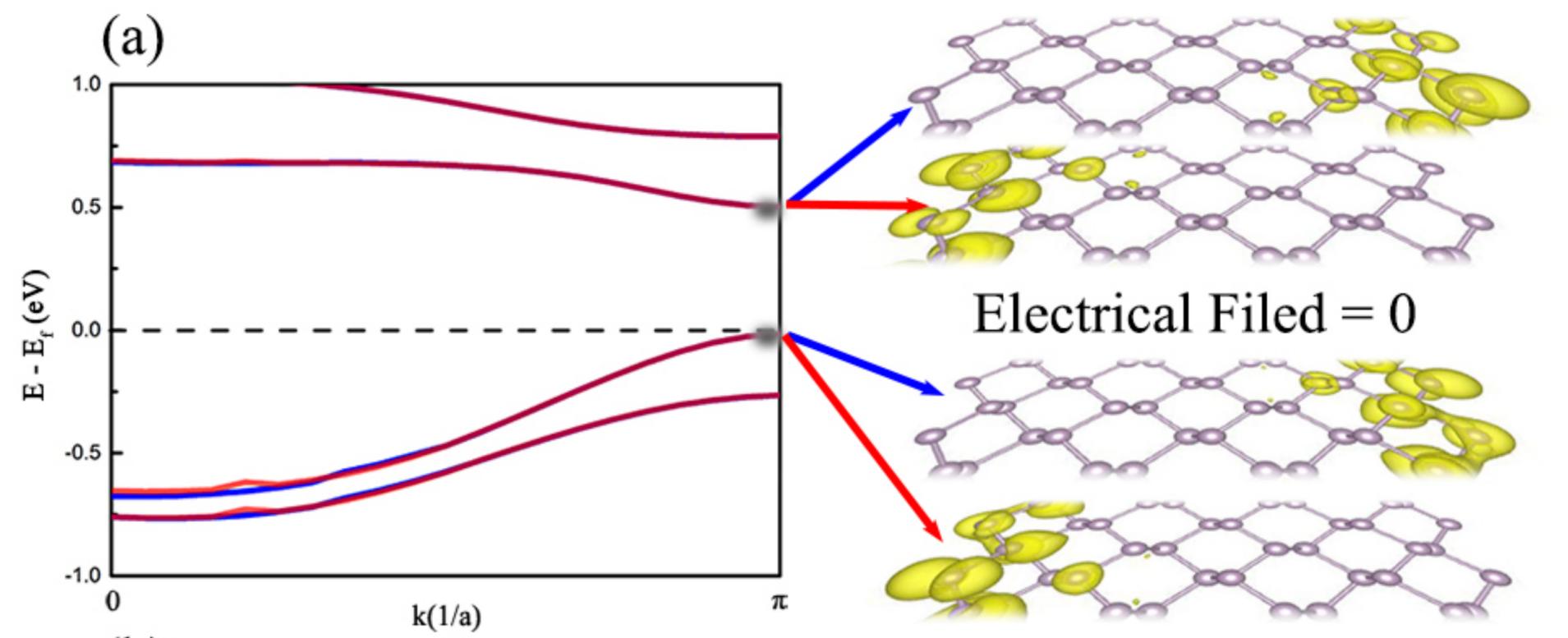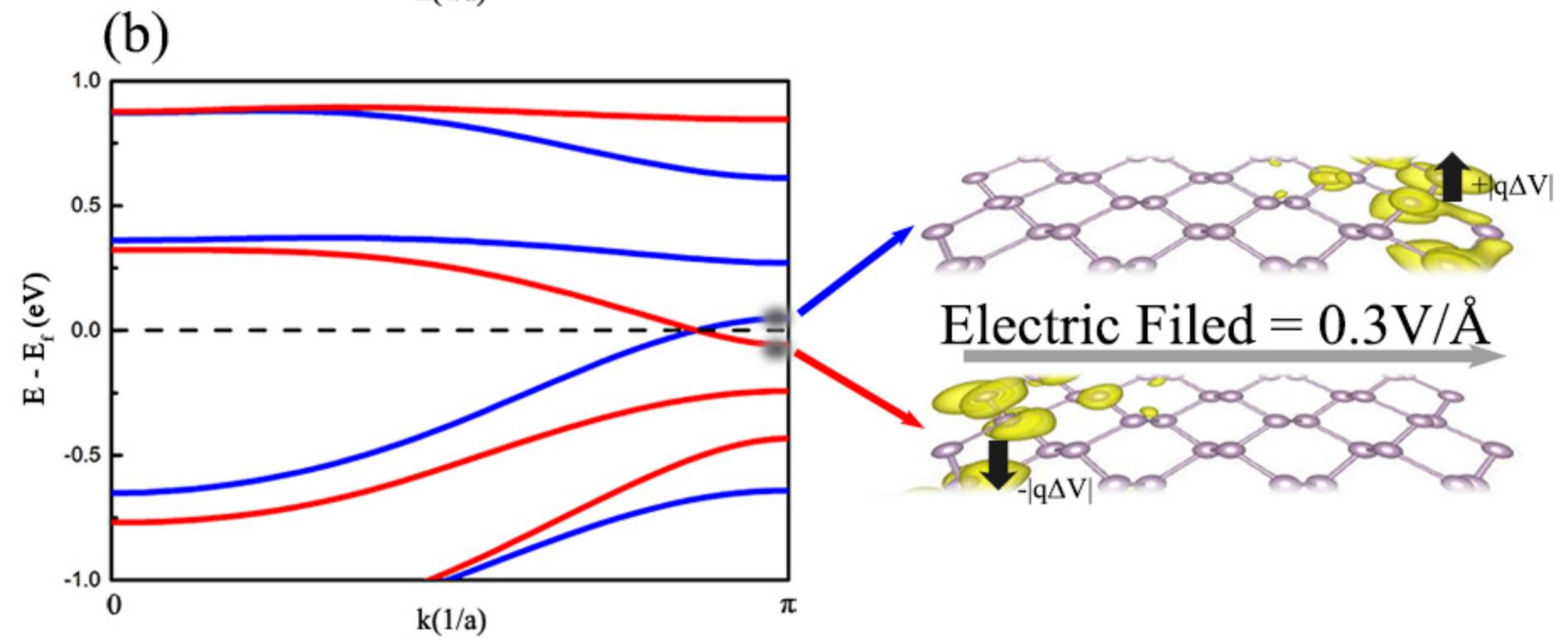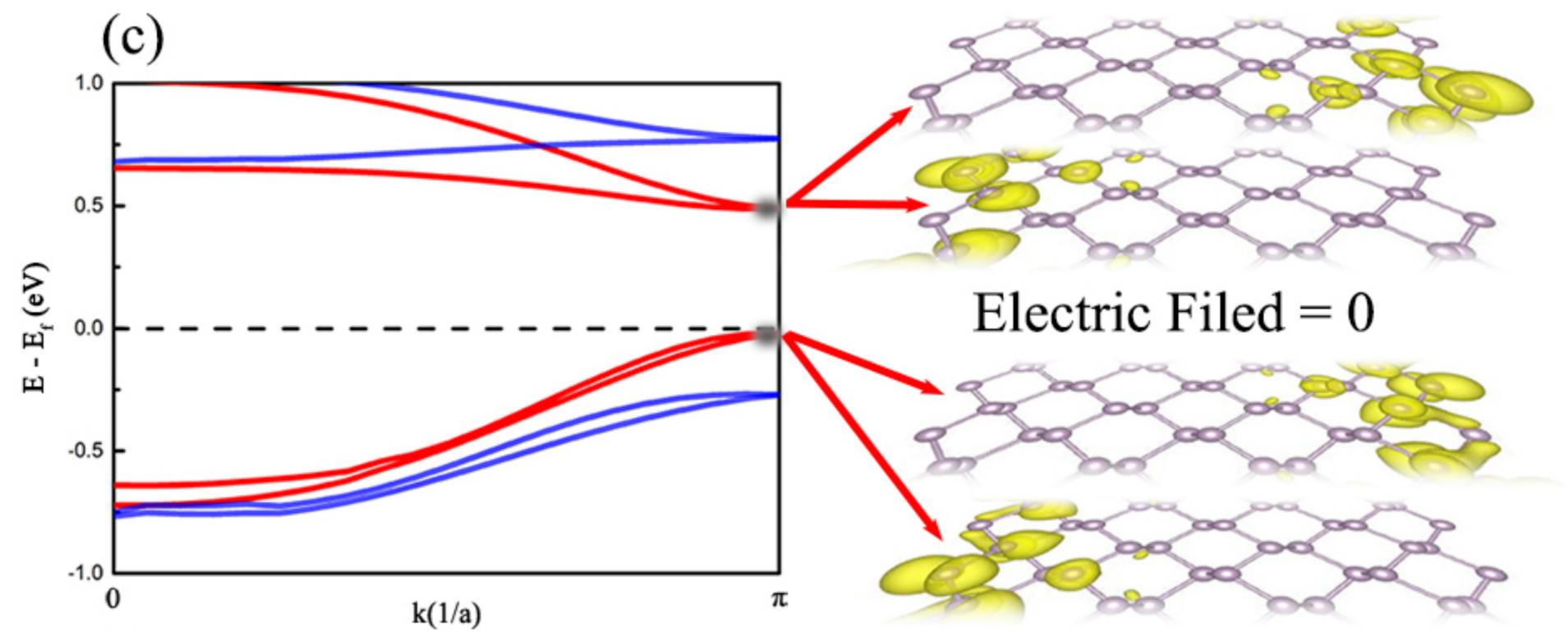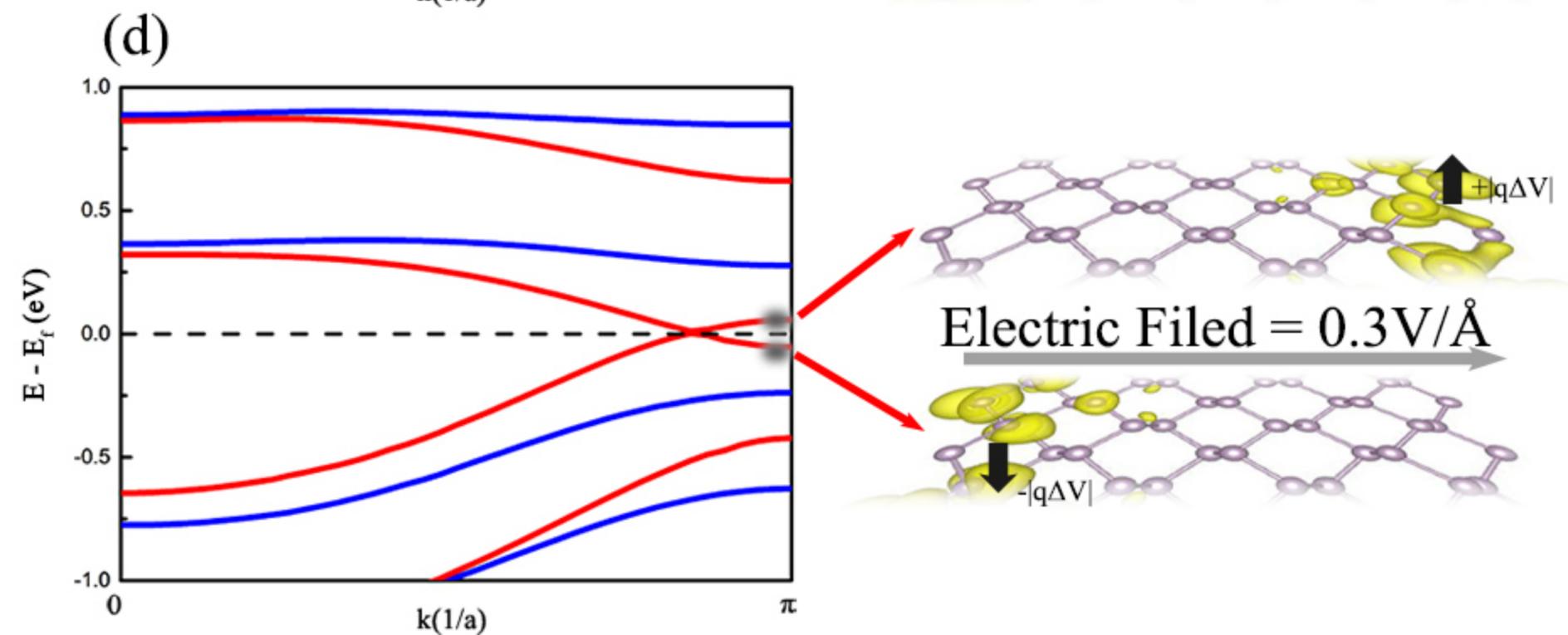

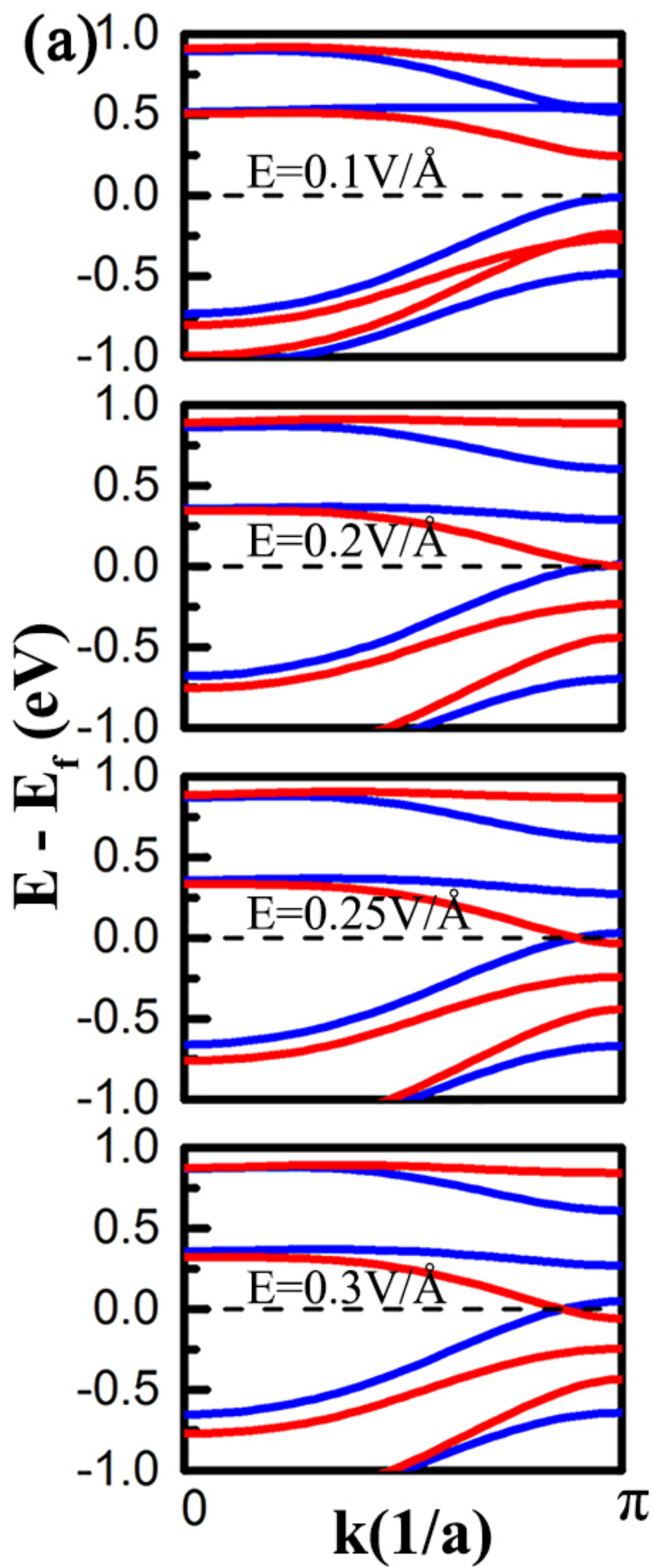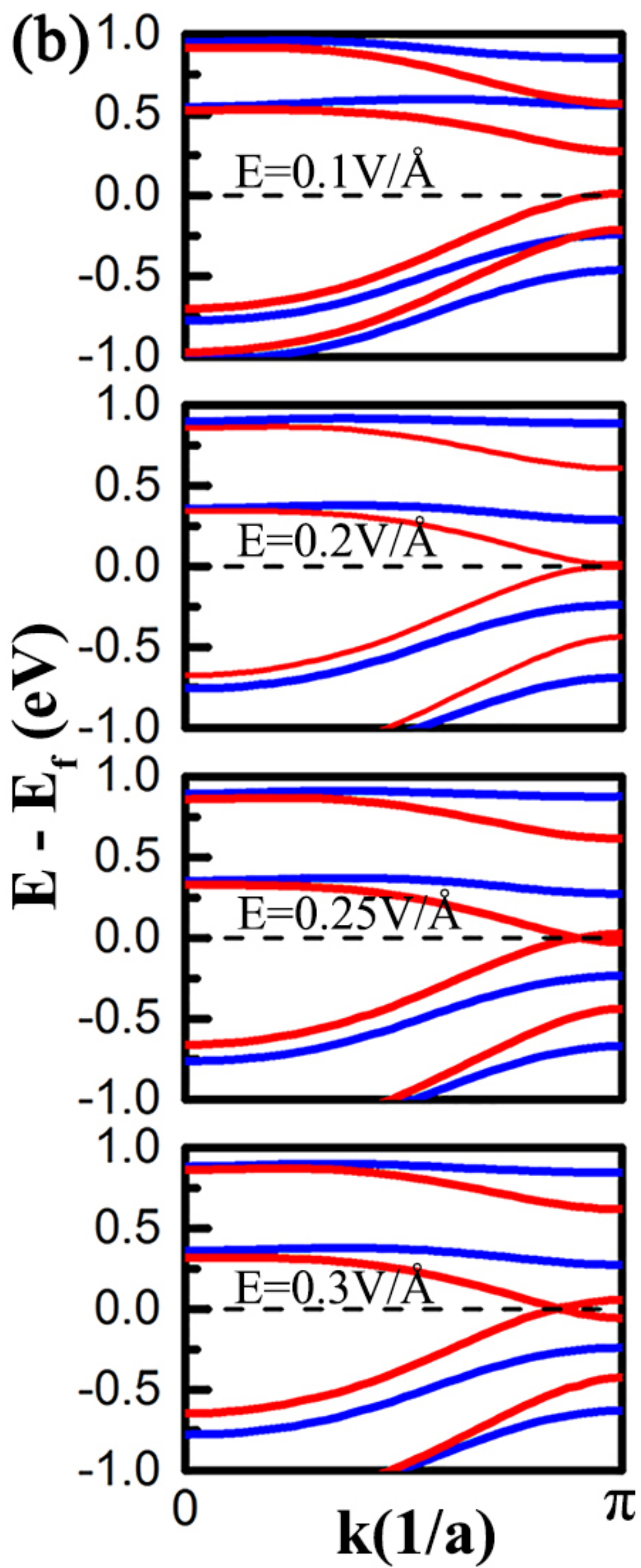

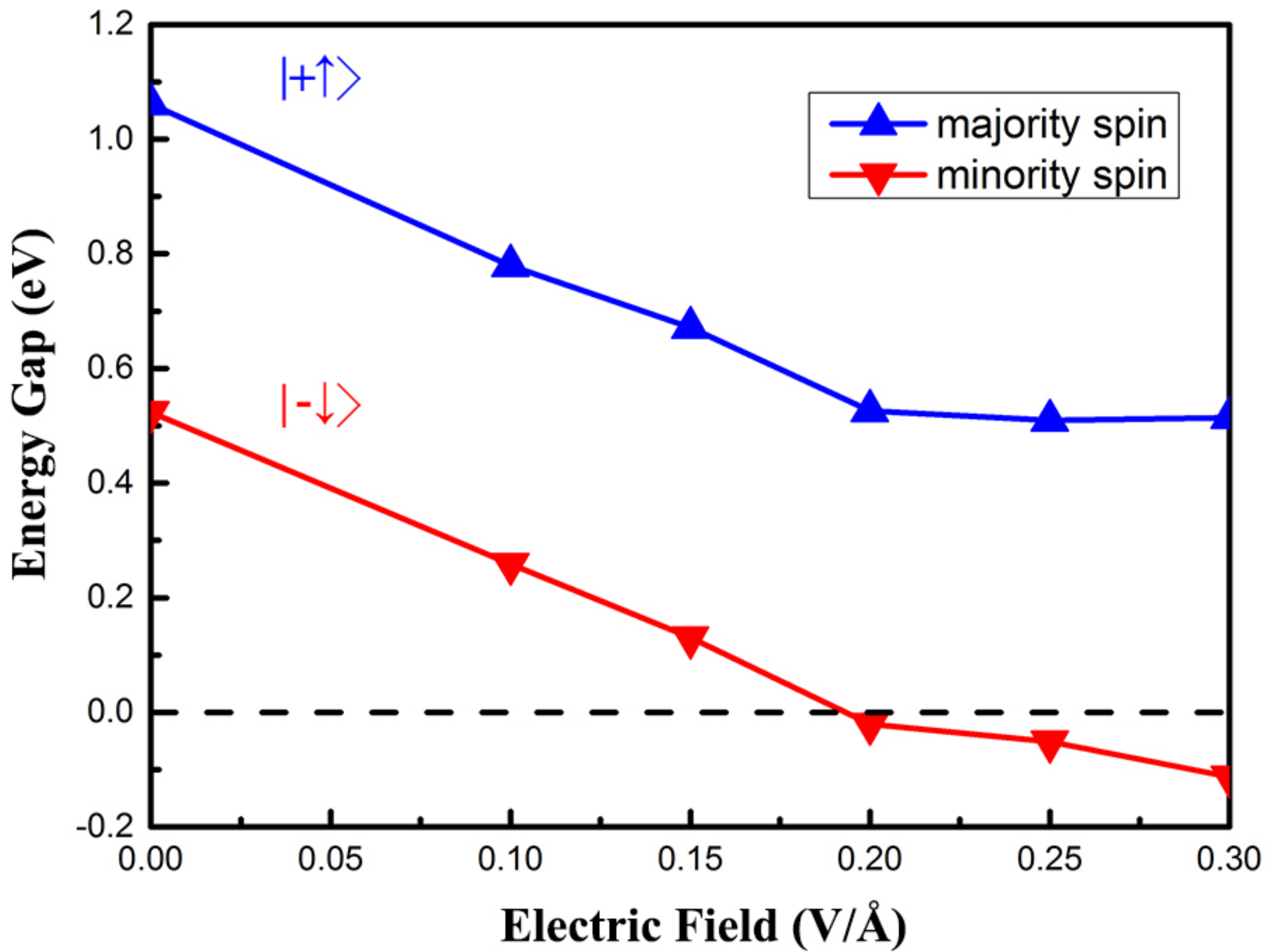